\documentclass[twocolumn,aps,altaffilletter]{revtex4} 
\usepackage{shortvrb} \let\|=| \MakeShortVerb\|
\usepackage{graphicx}
\usepackage{color}
\usepackage[colorlinks=true]{hyperref}
\usepackage{bm}
\usepackage{xspace}
\usepackage{times}

\newcommand{\YBCO}{YBa$_2$Cu$_3$O$_{7-x}$\xspace}
\newcommand{\units}[1]{\ensuremath{\,{\rm #1}}}

\begin{document}

\title{High performance magnetic field sensor based on Superconducting Quantum Interference Filters}

\author{P.~Caputo}%
\email{p.caputo@uni-tuebingen.de}%
\author{J.~Oppenl{\"a}nder}%
\author{Ch.~H{\"a}ussler}%
\author{J.~Tomes}%
\author{A.~Friesch}%
\author{T.~Tr\"auble}%
\author{N.~Schopohl}%
\affiliation{%
Lehrstuhl f\"ur Theoretische Festk{\"o}rperphysik, Universit{\"a}t T{\"u}bingen\\ 
Auf der Morgenstelle 14, 72076 T{\"u}bingen (Germany)
}%

\date{\today}

\begin{abstract}

  We have developed an absolute magnetic field sensor using Superconducting Quantum Interference Filter (SQIF) made of high-$T_c$ grain boundary Josephson junctions. The device shows the typical magnetic field dependent voltage response $V(B)$, which is sharp delta-like dip in the vicinity of zero magnetic field. When the SQIF is cooled with magnetic shield, and then the shield is removed, the presence of the ambient magnetic field induces a shift of the dip position from $B_0\approx 0$ to a value $B\approx B_1$, which is about the average value of 
  the earth magnetic field, at our latitude.
  When the SQIF is cooled in the ambient field without shielding, the dip is first found at $B\approx B_1$, and the further shielding of the SQIF results in a shift of the dip towards $B_0\approx 0$. The low hysteresis observed in the sequence of experiments (less than 5\% of $B_1$) makes SQIFs suitable for high precision measurements of the absolute magnetic field. The experimental results are discussed in view of potential applications of high-$T_c$ SQIFs in magnetometry.

\end{abstract}


\maketitle


For the last decades there has been much progress in the development of 
Superconducting Quantum Interference Devices (SQUIDs) as high performance magnetic field sensors (e.g. see \cite{Koelle_Review, Weinstock_96} and Refs. therein). Since they combine a high resolution with a relatively large bandwidth, SQUIDs have found a large variety of applications in magnetometry. SQUIDs can reach magnetic field sensitivity of few ${\rm fT/\sqrt{Hz}}$ when operated with on chip pick-up loops and flux focusing areas; ultra low magnetic field noise at low frequency is achieved by active field compensation electronics\cite{Forgacs} and bias-reversal techniques\cite{Koch_1983}. Sophisticated bio-magnetic instrumentation containing arrays of several hundreds of SQUID sensors are currently employed for non-invasive monitoring of brain activity, pre-surgical mappings, and bio-chemical investigations. SQUID applications are also used in geophysics to map the Earth's magnetic field and its gradients, and for long term monitoring of anomalous environmental magnetic field variations\cite{Machitani}. 
Although the present state of art in SQUID applications is quite impressive, in the last few years a number of novel devices has been proposed as alternative to SQUID magnetometers. Atomic magnetometers based on quantum coherence of magnetic levels in atoms, have been demonstrated to have an extremely high magnetic field sensitivity ($<1\units{fT/\sqrt{Hz}}$) in a reasonable bandwidth ($<30\units{Hz}$)\cite{Kominis_Nature2003}. In similar systems, magnetic field sensitivities comparable to the best SQUID systems have been reported in Refs.\cite{Bison_APB}. 

Recently, 
Superconducting Quantum Interference Filters (SQIFs) have been proposed as high resolution magnetometers \cite{Oppenlander:PRB,Haussler:JAP2001}. SQIFs are arrays of Josephson junctions (JJs) with a specially selected distribution of the loop areas, such that the magnetic field dependent voltage response $V(B)$ is a delta-like dip in the vicinity of zero magnetic field. 
The amplitude of the dip increases with the number of loops in the SQIF connected in series\cite{Oppenlander:APL_2003}.
The uniqueness of the dip and the high transfer factor (gain) $G\equiv\max_B\frac{\partial V}{\partial B}$ make the SQIFs attractive as absolute magnetic field sensors, even in unshielded environments, or as broad band 
amplifiers. 
Based on the interferometric principle of all loops, SQIFs are expected to have  a ultra-high sensitivity to the magnetic field, with a flux noise level of few $\units{fT/\sqrt{Hz}}$\cite{Oppenlander:PhysicaC-2002}. Thus, the intrinsically high dynamic range and the linearity in the voltage output represent the main advantages of using SQIFs. 

In this letter, we present our recent experiments on one dimensional (1D) SQIF magnetometers consisting of high-${T_c}$ Josephson junctions. All experiments are done in active micro-coolers, which do not lead to any degradation of the SQIF performance. When the SQIF is cooled with a magnetic shield, and then the shield is removed, the presence of the ambient magnetic field induces a shift of the dip position from $B_0\approx 0$ to a value $B\approx B_1$, which is about the average value of the earth magnetic field. Replacing the magnetic shield, the dip returns to its original value $B_0$. We have found that the two values are reproducible within the accuracy of our experimental set-up. 
We have also characterized the noise spectral density of the samples. The experimental results might be relevant for the development of SQIFs as magnetic field sensors.

%
\begin{figure}[!b]
  \centering
  \includegraphics*[width=5cm]{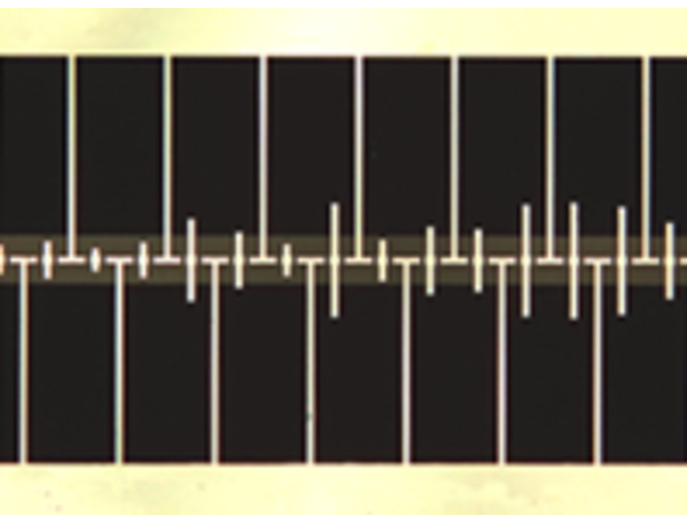}\\
  \includegraphics*[width=\columnwidth]{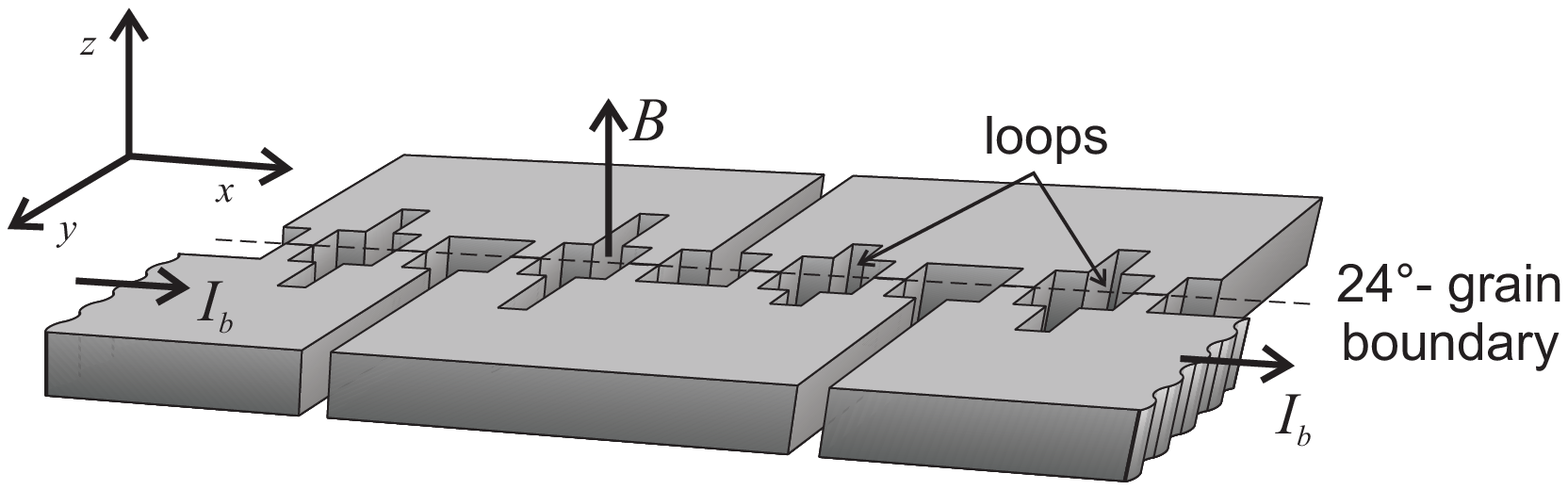}
  \caption{
    (a) Optical micrograph of a fragment of the serial 1D-SQIF. The typical distribution of the loop areas is clearly visible. (b) Sketch of the sample showing, not in scale, the loops, the electrodes, and the line along the grain boundary junctions; the direction of the magnetic field $B$ and of the bias current $I_b$ is indicated by arrows.
  }
  \label{sketches}
\end{figure} 
  \begin{figure*}[!htb]
    \centering
    \includegraphics*{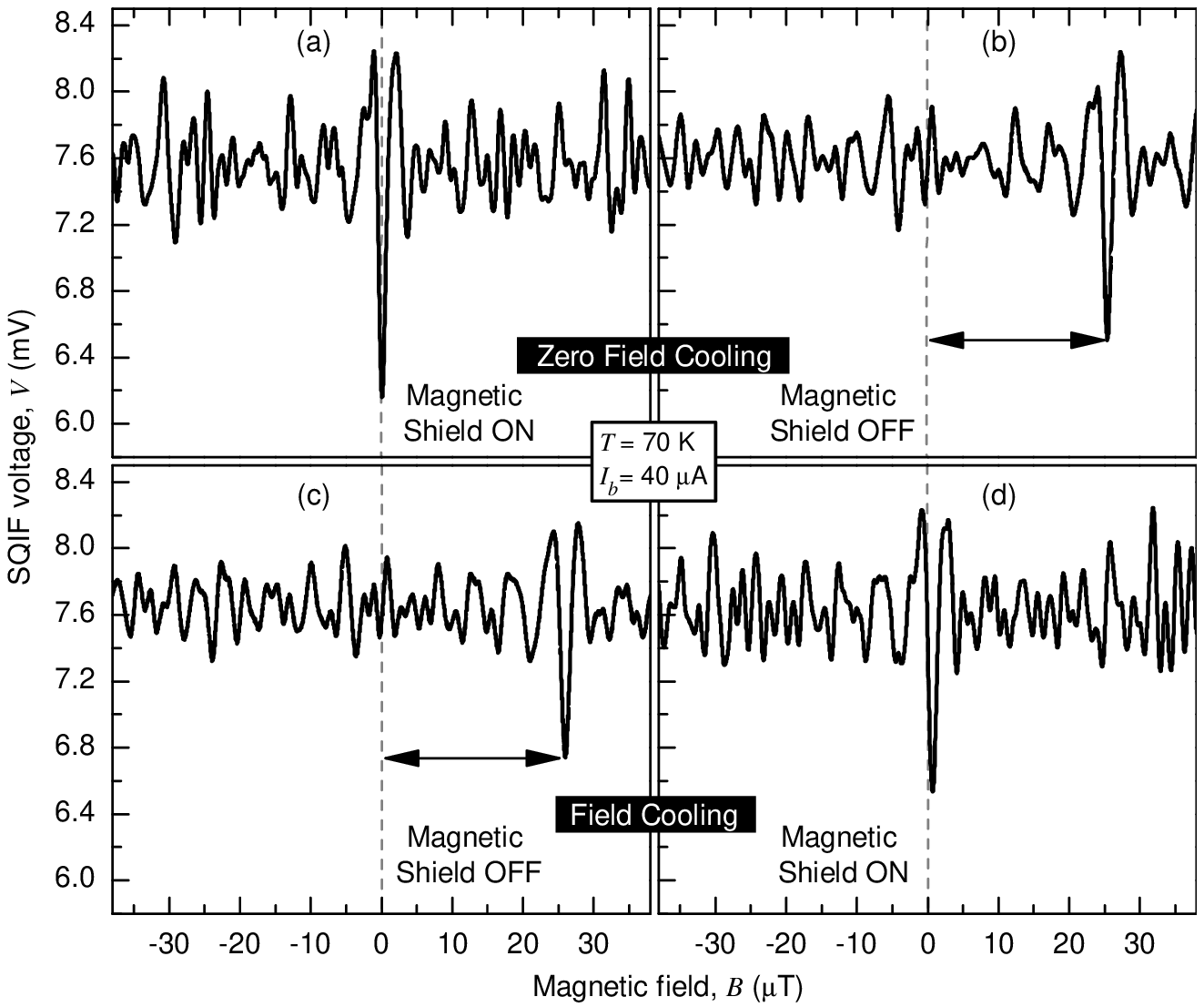}
         \caption{
      Set of experiments performed in zero field cooling (ZFC) and field cooling (FC) conditions. ZFC: $V(B)$ dependences measured with (a) and without (b) the magnetic shield; FC: $V(B)$ dependences measured without (c) and with (d) the magnetic shield. All curves were measured at $I_b=40\units{\mu A}$ and $T=70\units{K}$.
    }
    \label{dips}
  \end{figure*} 
 %
 %

The 1D-SQIFs are made of $N=100$ loops, which are placed along the grain boundary and connected in series. The voltage response is measured across the whole SQIF. The magnetic field is applied by means of a calibrated coil in the direction perpendicular to the loops. A fragment of the 1D-SQIF, with typical distribution of the loop areas, is shown in Fig.\,\ref{sketches}(a). The SQIFs are made of \YBCO grain boundary junctions, grown on 24$^{\circ}$-oriented bycristal substrates\cite{IPHT_Jena}. The grain boundary of the substrate runs along the dashed line shown in Fig.\,\ref{sketches}(b). The \YBCO electrodes [the black areas in Fig.\,\ref{sketches}(a)] have thickness of $200\units{nm}$, and a critical temperature of $82\units{K}$. 
The width of each junction is designed to be $1\units{\mu m}$; the distribution of loop areas ranges between $38 \ldots 224 \units{\mu m^2}$, over a base sequence of 10 loops, repeated 10 times to get 100 loops in total. In order to enhance the magnetic field sensitivity, the superconducting areas around the loops are extended over $100\units{\mu m}$. For a junction critical current $I_c$ of $12\,\mathrm{\mu A}$, the values of the parameter $\beta_L={L I_c}/{\Phi_0}$ range between $0.5 \ldots 1$. The junctions are intrinsically overdamped, and the typical $I_cR_n$ product is about $100 \,\mathrm{\mu V}$ (at $T=60 \,\mathrm{K}$). 
The area of the whole device is about $1.2\, \mathrm{mm^2}$. When required by the experiment, ambient magnetic fields were shielded by a room temperature mu-metal can. 
Samples were operated in Stirling micro-coolers\cite{AIM}, in a temperature range from $50$ to $80\,\mathrm{K}$. The use of the micro-cooler offers the possibility to have temperature stability ($\pm 0.1 \mathrm{K}$) over a wide temperature range, quick thermal cycles, and a full portability of the experiment. Although the micro-cooler is an active system (the cryogenic parts are driven by a 55 Hz compressor), we have    
performed low-noise measurements, and characterized the direct spectral noise density of the samples at $T=70 \,\mathrm{K}$, without using flux-locked loop electronics and/or bias reversal compensation techniques.

The SQIF top performance data correspond to a maximum voltage span of about $2.1\, \mathrm{mV}$ at a bias current $I_b=40\, \mathrm{\mu A}$ and a temperature $T=70\, \mathrm{K}$, at which the SQIF critical current is $I_c=30\units{\mu A}$. The dip is symmetric, and the side modulation has a maximum peak to peak amplitude of about $1\units{mV}$. The width of the dip is $\Delta B=3.3\, \mathrm{\mu T}$, and the maximum transfer factor is $G=3.5\, \mathrm{\mu V/nT}$. The dynamical range of the SQIF is as large as $30\, \mathrm{\mu T}$. The specially designed electrode symmetry prevents undesired shifts of the dip due to self field effects, and even at large bias currents (up to $2.5 I_c$) the position of the dip stays at the same magnetic field value. 

First, we made experiments with the chip mounted so that the loop plane is parallel to the earth surface.
Figures\,\ref{dips}(a--d) show a typical sequence of $V(B)$ curves measured with different cooling procedures. All curves refer to $I_b=40\, \mathrm{\mu A}$ and $T=70\, \mathrm {K}$.  First, we have performed a Zero Field Cooling (ZFC) [Fig.\,\ref{dips}(a)]. The chip is cooled down below $T_c$ ($T=70\, \mathrm{K}$) in the absence of magnetic field, which is shielded by the room temperature mu-metal can. As there is always a negligible residual magnetic field inside the shield, we are not in an {\em ideal} ZFC condition, and the dip is located at  $|<|B_0|>|=0.15\, \mathrm{\mu T}$, rather than at zero (in an ideal ZFC transition, no vortices should be frozen in the SQIF, neither in the superconducting electrodes nor in the loops). 
When 
we remove the magnetic shield, the dip promptly shifts to the position $|<|B_1|>|=26.5\, \mathrm{\mu T}$ [Fig.\,\ref{dips}(b)]. This is because the ambient magnetic field penetrates in the loops, and a compensation field is needed to reach a net zero field state (required for the existence of the dip). 
Even in unshielded conditions, the dip is stable; only its amplitude reduces and the pattern of the side modulation is slightly modified. 
The two states are fully reversible: we can move from the first state (magnetic shield ON, dip positioned at $|<|B_0|>|=0.15\, \mathrm{\mu T}$), to the second state (magnetic shield OFF, dip positioned at $|<|B_1|>|=26.5\, \mathrm{\mu T}$), and vice versa, by slowly removing/replacing the magnetic shield.  
As long as we do not perform a new thermal cycle through $T_c$, the two states are reproducible within the accuracy of our experimental set-up ($<$1\%). If a new ZFC transition is made, the dip might be located at slightly different values. We have computed a standard deviation $\Delta {B_0}/|<|B_0|>| \approx \Delta B{_1}/|<|B_1|>| \approx 2\%$,  obtained from a statistical analysis of the data distribution in $25$ thermal cycles. 

We have also performed Field Cooling (FC) experiments. The chip is cooled down to $T=70\, \mathrm {K}$ without the mu-metal shield, i.e. in the presence of the ambient magnetic field, and then biased at $I_b=40\, \mathrm{\mu A}$. In this case, the dip is found at about $|<|B_1|>|=26.1\, \mathrm{\mu T}$ [Fig.\,\ref{dips}(c)]; i.e., in order to reach a net zero field inside the loops, it is required approximately the same compensation field as in the ZFC case. This means that the vortices trapped in the films do not contribute significantly to the total field distribution, for example due to a symmetric distribution.  
Replacing the magnetic shield brings the dip at $|<|B_0|>|=1.1\, \mathrm{\mu T}$ [Fig.\,\ref{dips}(d)]. As in ZFC case, the two states can be reversibly exchanged, by means of the magnetic shield. However, as the FC procedure unavoidably induces an arbitrary amount of trapped vortices in the circuitry, a standard deviation of 5\% is found in this case, over 25 FC runs. The observed hysteresis, although relatively little, is uncontrollable.

In order to evaluate the magnetic field vector, we have positioned the chip so that the plane of the chip is perpendicular to the earth surface. In the ZFC case, and with magnetic shield, the dip is found at $|<|B_0|>|=0.10\, \mathrm{\mu T}$. If the SQIF is aligned with the N-S direction (as measured by a compass), the dip remains in this position even when the shield is removed.
By rotating the chip, the dip deviates from the initial value, and it reaches the value of $|<|B_1|>|=19.5\, \mathrm{\mu T}$ when the sample is rotated of $90^{\circ}$ with respect to the initial position, i.e. when the SQIF plane is perpendicular to N-S direction. Thus, we get an estimate of the module of the magnetic field vector $\mid{\mathbf B}_1\mid\,=33.5\units{\mu T}$.  

We have characterized the noise performance of the SQIF magnetometer in the active micro-cooler.  
By using low-noise and wide band operational amplifiers, the output noise level of the preamplifier is as low as $0.4\,\mathrm{nV/\sqrt{Hz}}$, in a bandwidth of $200\,\mathrm{kHz}$, and the $1/f$ corner frequency is around $10\,\mathrm{Hz}$. 
The compressor disturbances result in a peak at 55 Hz, and few higher harmonics. The measured frequency independent voltage noise level (white noise) of the SQIF is in agreement with the expected thermal noise level\cite{Clarke_JLTP76} [$\propto \sqrt{{R_d^2}/{R_s}}$, where $R_d$ and $R_s$ denote the SQIF differential and static resistance, respectively]. The resulting magnetic flux noise level, measured at the point of maximum $G$, is about $4\,\mathrm{pT/\sqrt{Hz}}$. 
Much higher magnetic field resolution ($\approx 70\units{fT/\sqrt{Hz}}$) has been recently achieved in SQIFs specially designed with a pick-up loop \cite{Schultze, Tomes}.

The presented experiments were aimed to test the feasibility to use SQIFs as magnetometers. We have not optimized all experimental conditions needed for an accurate detection, such as a reliable control of the sample alignment by means of high precision goniometers, or the shielding of occasional  sources of  magnetic field in the vicinity the sample (unshielded instruments, etc.). 
The ZFC experiments have revealed high accuracy in determining the dip position 
and have allowed to give a fair estimation of the earth magnetic field vector.  
Thus, used in Zero Field Cooled conditions, SQIFs might serve as high accuracy absolute magnetometers. In addition, the measured flux noise sensitivity, although not optimized in the presented sample design, might allow to use SQIFs as detectors of weak magnetic signals, even at low frequency. An improvement of the device sensitivity will be realized by an on-chip pick-up coil.

\end{document}